\documentclass[aps,prd, reprint, showpacs,superscriptaddress,
floatfix,amsmath,amssymb]{revtex4-1}
\usepackage{lipsum}
\usepackage{multirow}
\usepackage{graphicx}
\usepackage{caption}
\usepackage{subcaption}
\usepackage{bm}
\usepackage{color}
\usepackage[table,xcdraw]{xcolor}

\def\baa#1\eaa{\begin{align*}#1\end{align*}}
\def\bs#1\es{\begin{split}#1\end{split}}

\newcommand{\f}[2]{\frac{#1}{#2}}

\newcommand{\be}{\begin{eqnarray}}
\newcommand{\ee}{\end{eqnarray}}
\newcommand{\bea}{\begin{eqnarray*}}
	\newcommand{\eea}{\end{eqnarray*}}
\def\ba#1\ea{\begin{align}#1\end{align}}
\def\bm#1\em{\begin{pmatrix}#1\end{pmatrix}}

\def \bit{\begin{itemize}\setlength\itemsep{0em}} 
\def \eit{\end{itemize}}

\usepackage{hyperref}

\begin{document}

\date{\today}

\title{Prospects for Measuring Planetary Spin and Frame-Dragging \\
in Spacecraft Timing Signals}

\author{Andreas Sch\"arer}
\email{andreas.schaerer@physik.uzh.ch}
\affiliation{Department of Physics, University of Zurich,
Winterthurerstrasse 190, 8057 Zurich, Switzerland}

\author{Ruxandra Bondarescu}
\email{ruxandra@physik.uzh.ch}
\email{ruxandrab7@gmail.com}
\affiliation{Department of Physics, University of Zurich,
Winterthurerstrasse 190, 8057 Zurich, Switzerland}

\author{Prasenjit Saha}
\affiliation{Department of Physics, University of Zurich,
Winterthurerstrasse 190, 8057 Zurich, Switzerland}
\affiliation{Institute for Computational Science, University of Zurich,
Winterthurerstrasse 190, 8057 Zurich, Switzerland}

\author{Raymond Ang\'elil}
\affiliation{Institute for Computational Science, University of Zurich,
Winterthurerstrasse 190, 8057 Zurich, Switzerland}

\author{Ravit Helled}
\affiliation{Institute for Computational Science, University of Zurich,
Winterthurerstrasse 190, 8057 Zurich, Switzerland}
\affiliation{Department of Geosciences, Tel Aviv University, Tel Aviv 69978
Israel}

\author{Philippe Jetzer}
\affiliation{Department of Physics, University of Zurich,
Winterthurerstrasse 190, 8057 Zurich, Switzerland}

\begin{abstract}
Satellite tracking involves sending electromagnetic signals to Earth. 
Both the orbit of the spacecraft and the electromagnetic signals themselves are affected by the curvature of spacetime.
The arrival time of the pulses is
compared to the ticks of local clocks to reconstruct the orbital path
of the satellite to high accuracy, and to implicitly measure general
relativistic effects.
In particular, Schwarzschild space curvature (static)
and frame-dragging (stationary) due to the planet's spin affect the satellite's
orbit. The dominant relativistic effect on the path of the signal photons is Shapiro delay due to static space curvature.
We compute these effects for some current
and proposed space missions, using a Hamiltonian formulation in four
dimensions.
For highly eccentric orbits, such as in the {\em Juno} mission and in the {\em Cassini} Grand Finale, the relativistic effects have a kick-like nature, which could be advantageous for detecting them if their signatures are properly modeled as functions of time.
Frame-dragging appears, in
principle, measurable by {\em Juno} and {\em Cassini}, though not by
{\em Galileo} $5$ and $6$.  Practical measurement would require disentangling
frame-dragging from the Newtonian ``foreground'' 
such as the gravitational quadrupole which has an impact on both the spacecraft's orbit and the signal propagation.
The foreground problem remains to be solved.
\end{abstract}

\maketitle

\section{Introduction}

General relativity (GR) describes gravitation as a consequence of a curved
four dimensional spacetime \cite{SSQ.Iorio.editorial,SSQ.Debono.Smoot}. In most astrophysical systems, however,
dynamics are dominated by Newtonian physics and GR only provides very
small perturbations.  Near a mass $M$, the relativistic perturbations
on an orbiting or passing body depend mostly on the pericenter
distance, which we call $p$, in units of the gravitational radius $GM/c^2$.
Newtonian effects are of order $O(p^{-1/2})$.  The largest
relativistic perturbation is time dilation, and is of
$O(p^{-1})$.  Space curvature, referring to space-space
terms in the metric tensor, enters dynamics at $O(p^{-3/2})$.  At
$O(p^{-2})$ mixed space-time metric terms enter the
dynamics; these correspond to frame-dragging effects, in which a
spinning mass drags spacetime in its vicinity and thereby affects the
orbit and orientation of objects in its gravitational field.
Gravitational radiation corresponds to dynamical effects of
$O(p^{-3})$. In post-Newtonian notation, X\thinspace PN (e.g. $1$\thinspace PN, $2$\thinspace PN, \dots)
corresponds to $O(p^{-{\rm X}-1/2})$.  In the Solar System, $p$ is
very large in gravitational terms: $\sim10^8$ or more.  In close
binary systems $p$ can be much less. In binary pulsars the
combination of comparatively low $p\sim10^5$ with the long-term
stability of pulsar timing enables the measurement of relativistic
effects down to gravitational
radiation\cite{SSQ.PulsarTaylor1994,SSQ.PulsarKramer2006}.

All the same effects are, in principle, present for artificial Earth
satellites, but since $p\sim10^9$, they are much weaker.  Nonetheless,
until now the frame-dragging effect of the Earth's spin has been
detected in two different ways: (1)~the LAGEOS and LARES satellites
used laser ranging to measure orbital perturbations from
frame-dragging \cite{SSQ.ciufolini2004confirmation,SSQ.ciufolini2016}
(some aspects are still controversial
\cite{SSQ.iorio2011phenomenology,SSQ.renzetti2013history,SSQ.Renzetti.Monte.Carlo,SSQ.Iorio.comment});
(2)~Gravity Probe~B measured the effects of frame-dragging on the
orientation of onboard gyroscopes \cite{SSQ.everitt2011gravity}.  GPS
satellites are well known to be sensitive to time dilation
\cite{SSQ.ashby2003} and upcoming missions will put even more precise
clocks in orbit. In the Atomic Clock Ensemble in Space (ACES) mission
\cite{SSQ.cacciapuoti2011atomic}, two atomic clocks will be brought to
the ISS in order to perform such experiments. However, the ISS is not
the optimal place to probe GR and a dedicated satellite on a highly
eccentric orbit would be desirable. Its proximity to Earth and high
velocity at pericenter would boost relativistic effects and therefore
improve the measurements. Several such satellites equipped with an
onboard atomic clock and a microwave or optical link on very eccentric
orbits, such as STE-QUEST, have been discussed and studied
\cite{SSQ.Altschul2015501}. Such missions would not only be very
interesting to probe gravity but also have a plethora of applications,
e.g., in geophysics \cite{SSQ.Bondarescu2012,SSQ.VolcanoPaper}.

Missions like {\it Juno} and {\it Cassini} present new possibilities
for measuring relativistic effects around the giant planets in our
Solar System. The basic idea goes back to the early days of general
relativity, when Lense and Thirring \cite{SSQ.lense-thirring1918}
showed that the orbital plane of a satellite precesses about the spin
axis of the planet ---that is what we now call frame-dragging--- and
identified the expected precession of Amalthea's orbit by $1'\,53''$
per century as the most interesting case. Recent work has drawn
attention to the corresponding precession in the case of {\em Juno}
\cite{SSQ.helled2011jupiter,SSQ.iorio2013possible,SSQ.iorio2010juno}
and other systems
\cite{2011PhRvD..84l4001I,2012GReGr..44..719I,SSQ.Iorio.Spin.Planets}.

The classical Lense-Thirring precession is an orbit-averaged effect.
This comes with the problem that the very small precession due to
relativity is masked by much larger non-relativistic precession,
making it very hard to identify the relativistic contribution. For
example, most of Mercury's observed precession is due to Newtonian
planetary perturbations, the relativistic contribution being only
about $7\%$ of the total \cite{SSQ.Park.Mercury}.
It is better to have something with a specific
time dependence that can be filtered out. 

Here, we extend the work of Angelil {\it et al.} for terrestrial satellites
\cite{SSQ.SpaceClox2014} and the Galactic center
\cite{SSQ.kannansaha2009,SSQ.2009ApJ...703.1743P,SSQ.angelilsaha2010,SSQ.angelilsahamerritt2010,SSQ.2011ApJ...734L..19A,SSQ.2014MNRAS.444.3780A,SSQ.2017ApJ...834..198Z}
and apply it to other planets in the Solar System. 
Since the orbits
are dominated by Newtonian physics, and relativity only contributes
very small perturbations, their investigation is numerically
challenging. In earlier work
\cite{SSQ.SpaceClox2014} the orbits were therefore simulated with smaller
semi-major axes compared to the real orbit and then, by knowing how
the individual effects scale, the redshift curves were obtained by
correctly scaling up. Here, we use an arbitrary precision code
instead.

We look at an idealized model where a spacecraft sends electromagnetic signals to a ground station. Comparing the relativistic $4$-momentum of the emitted photon to that of the one received at the station allows determining a redshift $z$ (see Eq. \eqref{equ redshift}).
Equivalently, one can consider an orbiting clock which sends out signals corresponding to the ticks of the clock \cite{SSQ.angelilsaha2010,SSQ.SpaceClox2014}.
Then, the redshift arises when two photons emitted by the spacecraft at an interval
of proper time $\Delta \tau$ travel through curved spacetime hitting
the observer with a difference in the arrival time $\Delta t = \Delta
\tau (1 + z)$.
In both cases, a one-way signal transfer is considered.
Typically, satellite communication systems allow two-way signal transfer.
For a comparison of distant ground clocks like done with ACES, this leads to a first order cancellation of the position errors of the clocks \cite{Duchayne}.

To estimate the relativistic effects, we solve for the trajectory of
\begin{enumerate}
\item the satellite in a curved spacetime, and 
\item the photons (or propagating ticks from the frequency standard) as they propagate to the receiving station
\end{enumerate}
in a given gravitational field.
Both the satellite and the photons follow geodesics of the
metric and can be obtained by integrating the relativistic Hamiltonian,
expanded in velocity orders.
The redshift depends on both the classical Doppler shift as well as a number of relativistic effects.
Both trajectories are generated numerically via a
simulation code that handles multiple scales through variable
precision. The effects are modulated by
the varying gravitational field.

The paper proceeds as follows: Sec. \ref{SSQ.GR} describes the approximations we make for the spacetime outside a planet. It presents the Hamiltonian system that is being solved numerically with the higher order relativistic effects, and their respective scalings with orbital size. We then compute the magnitude of the spin parameter, of Schwarzschild precession and frame-dragging effects for the planets in our 
Solar System, and report them relative to the effects around Earth for orbits
of similar proportionality. Sec. \ref{SSQ.sec:missions} A and B apply this formalism to the {\it Juno} and {\it Cassini} Missions. Sec. \ref{SSQ.sec:missions} C discusses the {\em Galileo} 5 and 6 satellites and other proposed Earth-bound missions. In particular, it discusses the importance of eccentricity in detecting relativistic effects.


Conclusions and potential future directions are presented in Sec. \ref{SSQ.conclusions}.


\section{General Relativistic effects}
\label{SSQ.GR}

Calculating relativistic effects fundamentally involves two things:
the metric and the geodesic equations. The well-known epigram by
J.A. Wheeler states {\it Spacetime tells matter how to move, matter tells
  spacetime how to curve.}  The metric is known explicitly in terms of
the masses, including mass multipoles, and spin rates.
The geodesic equation, in general, requires a numerical solution.
However, in special or approximate cases analytical solutions also exist \cite{SSQ.Klioner.Kopeikin.1992,SSQ.Ashby.Bertotti.2010,SSQ.Hees.2014,SSQ.Crosta.2015,DOrazio}.

We wish to understand how different terms in the metric, in particular the spin part, affect the observable redshift signal.  To do this, we will numerically
integrate the geodesic equations with different metric terms turned on
and off and compare the resulting redshift signal curves.

In Sec. \ref{subsec Basic formulation} we briefly introduce the Hamiltonian formalism and the formula for calculating the redshift.
This is followed by Sec. \ref{subsec The expanded Hamiltonian}, which discusses the expansion of both the orbital as well as the signal Hamiltonian.
In Sec. \ref{subsec The spin parameter} we discuss the spin parameter and in \ref{subsec Keplerian elements} we discuss the cumulative changes of the Keplerian elements due to orbital relativistic effects. Finally, in Sec. \ref{subsec scaling} we investigate how the sizes of the relativistic signals scale for the different planets in the Solar System.

\subsection{Basic formulation}
\label{subsec Basic formulation}

We work with the geodesic equations in four dimensions, in Hamiltonian
form.  The independent variable is not time, but the affine parameter,
which is just the proper time in arbitrary units.  Although the
formalism seems complex, it actually tends to lead to simpler
equations \cite{SSQ.angelilsaha2010,SSQ.SpaceClox2014} than other
formulations.

For any spacetime metric, the geodesic equations may be expressed in
Hamiltonian form as
\ba
\frac{dx^\mu}{d\lambda} =   \frac{\partial H}{\partial p_\mu} \qquad
\frac{dp_\mu}{d\lambda} = - \frac{\partial H}{\partial x^\mu}
\ea
where
\ba
H = {\textstyle\frac12} g^{\mu\nu}(x^\alpha) p_\mu p_\nu
\ea
with $x^\mu = (t, r^i)$ being the four-dimensional coordinates,
$p_\mu = (p_t, p_i)$ being the canonical momenta, and $\lambda$
being the affine parameter.



The satellite at position $\vec{r} = (r^i)$ orbiting with 4-velocity $u^\mu_\text{emit}$ emits a photon with 4-momentum $p_\mu^\text{emit}$ which arrives at an observer (having velocity $u^\nu_\text{obs}$) with momentum $p_\nu^\text{obs}$. The redshift is then given by
\ba
\label{equ redshift}
z = \frac{p_\mu^\text{emit} u^\mu_\text{emit}}{p_\nu^\text{obs} u^\nu_\text{obs}} - 1.
\ea
For a distant observer at rest, the redshift for orbital effects reduces to
\ba
z = \f1c u^t_\text{emit} - \f1c u^\text{LOS}_\text{emit} - 1,
\ea
where $u^\text{LOS}_\text{emit}$ is the satellite's velocity along the line of sight.

\subsection{The expanded Hamiltonian}
\label{subsec The expanded Hamiltonian}

In this subsection we use geometrized units. That is, $\vec r$ is
measured in units of $GM/c^2$ where $M$ is the planetary mass, while
$t$ is measured in units of $GM/c^3$.  The momentum is dimensionless.
Since the orbits considered are close to Keplerian, the
order-of-magnitude relations
\ba \label{SSQ.vc-orders}
|\vec p| \sim \frac vc, \qquad r \sim \left(\frac vc\right)^{-2}
\ea
will hold, where $v$ is the orbital speed.
The time-momentum $p_t$ is constant and its value only affects internal units of a
calculation.  It is convenient to set $p_t=-1$.

As usual in post-Newtonian celestial mechanics, we order contributions
in powers of $v/c$.  These correspond to different physical effects.
Moreover, the ordering in powers of $v/c$ is different for the
spacecraft orbit and the light signals.  Accordingly, we consider two
Hamiltonians, as follows.
\begin{equation}
\begin{aligned}
H^\text{orbit} &= H^\text{equiv-prin} + H^\text{Schwarzschild}
                + H^\text{spin} \\
H^\text{signal} &= H^\text{Minkowski} + H^\text{Shapiro}
\end{aligned}
\end{equation}
Since there is only one spacetime, $H^\text{orbit}$ and
$H^\text{signal}$ are just different approximations to the same
underlying Hamiltonian.

The orbit of the satellite is dominated by
\ba
\label{SSQ.equ Hamiltonian TD}
H^\text{equiv-prin} = - \frac{p_t^2}{2} + \left( - p_t^2 \,
U(\vec{r}\,) + \frac{{\vec{p}\,}^2}{2}\right)
\ea
where $U(\vec r)$ is minus the Newtonian gravitational potential, to
leading order $1/r$ but also including multipole moments $J_n$ as well
as the tidal potential due to the Sun and other planets.  The first
term on the right is of order unity, while the bracketed part is of
order $v^2/c^2$.  This Hamiltonian leads to a Newtonian orbit and
redshift contribution of order $v/c$, together with a time dilation
effect of order $v^2/c^2$.  Gravitational time dilation is a basic
consequence of the geometric description of spacetime, i.e., the
principle of equivalence. Indeed, equation~(\ref{SSQ.equ Hamiltonian TD})
is the simplest Hamiltonian consistent with the equivalence principle that
gives the correct Newtonian limit.  Moving clocks tick slower than
stationary ones. So do clocks in a gravitational field. For an
orbiting clock, both effects are equal to leading order.  The ground
station will have its own time dilation too, of course, and the
difference is what matters. Time dilation causes the localization of a
satellite to be off by kilometers, which has already been taken into
account by the early phases of GPS. While this relativistic effect is well
established, the {\em Galileo} satellites will measure it to unprecedented
precision.

Since higher order relativistic effects cause small changes in the
redshift, they can be studied perturbatively. We investigate each
effect individually by adding it to $H^\text{equiv-prin}$, and
computing the cumulative redshift. The redshift perturbation is
obtained by subtracting the redshift when the effect is artificially
turned off.

The next contribution to $H^\text{orbit}$ is
\ba \label{SSQ.Hschw}
H^\text{Schwarzschild} = - \frac{p_t^2}{r^2} - \frac{{\vec{p}\,}^2}{r}
\ea
which introduces the effect of space curvature in the Schwarzschild
spacetime.  It is easy to verify from equation (\ref{SSQ.vc-orders}) that the
Hamiltonian terms are of order $s\,v^4/c^4$, and they contribute to
redshift at order $s\,v^3/c^3$, where $s$ is the spin parameter. Note that the $s$ is larger
for planets ($\sim 10^2-10^3$) than for more compact systems like black holes ($s\sim 1$) and thus 
the spin terms are significantly larger than what one would expect from just looking at velocity order.

The leading-order frame-dragging effect arises when adding the term
\ba
H^\text{spin} \label{SSQ.Hspin}
= - \frac{2 p_t}{r^3} \, \vec{p} \cdot (\vec{s}\times\vec{r}) .
\ea
This term is of order $s\,v^5/c^5$ and contributes a redshift effect of
order $s\,v^4/c^4$.  Frame-dragging is due to the rotation of the central
mass, which spins with $\vec{s}$, and depends linearly on the spin
parameter $s = |\vec{s}|$.  At next higher order, the dominant term is
a spin-squared term, i.e., it is proportional to $s^2$
\cite{SSQ.SpaceClox2014}.  This effect has never been measured before.
But since $s$ is quite large for planets (see Table \ref{SSQ.tab: Spin}),
probing this effect should be within the scope of future satellite
missions.

The leading multipole contribution comes from $J_2$ in the Newtonian Hamiltonian \eqref{SSQ.equ Hamiltonian TD} and scales as $1/r^3$. 
Therefore, it has a different $r$-dependence as the relativistic effects discussed here.
The relativistic effect with the same $r$-scaling would be the spin-squared effect.

The main contribution to the redshift comes from the velocity along
the line of sight.  Therefore, in order to measure a certain
relativistic effect, it is desirable to have an
orbit-observer-configuration where the relativistic effect has a
significant contribution to the line of sight velocity.  For first
order spin, the leading contribution is given by
\ba
\label{SSQ.equ: Delta z Spin}
\Delta z_\text{spin}
= - \f{2}{r^2} \vec{s} \cdot (\hat{r}\times\hat{b}),
\ea
where $\hat{b}$ is the unit vector pointing from the satellite towards
the observer.  Interestingly, the spin related redshift contribution has no explicit
dependence on the satellite's velocity.

The signal photons travel to leading order on a straight line. The
leading relativistic effect, leading to a slight bending, is Shapiro
delay.  This part is best analyzed after transforming to a Solar System
frame.  The signal Hamiltonian is given by the sum of
\ba
H^\text{Minkowski} = - \f{p_t^2}{2} + \f{{\vec{p}\,}^2}{2}
\ea
and
\ba
H^\text{Shapiro} = - U({\vec p}\,) \left( p_t^2 + {\vec{p}\,}^2 \right) \,.
\ea
At the next order of expansion, further Shapiro-like terms as well as 
spin terms appear. However, they are expected to be too small to be
measured.
The effect of frame-dragging on light signals was calculated, e.g., by \cite{SSQ.Kopeikin.Spin.Light,SSQ.Wex.Kopeikin.Spin.Pulsar}.

\subsection{The spin parameter}
\label{subsec The spin parameter}

The dimensionless spin parameter of a celestial body is given by
\begin{equation}
s = \frac{c}{GM^2} \int \rho(\vec{x})\,\omega(\vec{x})\,r_\perp^2\,d^3\vec{x} \,.
\end{equation}
For solid-body rotation ($\omega=2\pi/P$, where $P$ is the spin period) the above
expression reduces to
\begin{equation}
s = 2\pi \times \hbox{MoI} \times \frac{c}{gP}
\end{equation}
where
\begin{equation}
\hbox{MoI} = \frac{1}{MR^2} \int \rho(\vec{x})\,r_\perp^2\,d^3\vec{x}
\end{equation}
is the dimensionless moment of inertia and $g=GM/R^2$ is the surface
gravity, where $R$ is the average radius of the body.
For realistic density and
$\omega$ profiles
\begin{equation} \label{simple-spin}
s \sim \frac{c}{gP}
\end{equation}
is still a useful rough estimate.  It may be convenient to remember it
as the number of days needed to reach the speed of light from an
acceleration of one $g$.

For yet another interpretation of the spin parameter, let us consider
two speeds: the surface speed of a spinning planet $v_s\sim R/P$ and
the launching speed needed to send something into orbit from the
surface $v_l^2\sim gR$.  In terms of these speeds, the approximate
formula \eqref{simple-spin} becomes
\begin{equation} \label{simple-spin2}
s \sim \frac{cv_s}{v_l^2} \raise.5ex\hbox{.}
\end{equation}
The maximal-spinning situation $v_s\approx v_l$ corresponds to a
planet spinning so fast that it almost breaks up under centrifugal
forces.  In this limit $s\sim c/v_l$.  Recalling the orders in
$H^\text{spin}$ in equation \eqref{SSQ.Hspin}, we can see that that
Hamiltonian term would be of order $v^4/c^4$ and the corresponding
redshift effect would be of order $v^3/c^3$.  That is, for a
low-orbiting spacecraft above a maximally-spinning planet,
relativistic spin effects will be comparable in size to
space-curvature effects.

\subsection{Keplerian elements}
\label{subsec Keplerian elements}

A Keplerian orbit is described by the Keplerian elements $a,e,\Omega,I$ and $\omega$. While $a$ and $e$ describe the size and the eccentricity of the ellipse, the three angles describe its orientation with respect to some reference plane.

For a relativistic orbit this is not true anymore, as the relativistic
effects induce deviations from Keplerian motion.  In principle,
however, it is still possible to determine the instantaneous Keplerian
elements at each point along the orbit: These correspond to a
Keplerian orbit having exactly the same velocity as the relativistic
one at a given position.

It is well-known that space curvature leads to a precession of the pericenter
\ba
\Delta \omega_\text{SS} = \frac{G M}{c^2} \frac{6\pi}{a(1-e^2)}
\label{SSQ.equ perihelion precession schwarzschild}
\ea
for one orbit.

However, $\omega$ is not shifted evenly along the orbit, in fact, there is almost no shift during most of the orbit, but around pericenter there is a kick-like shift.
Similarly, there is a precession of the pericenter due to frame-dragging \cite{SSQ.lense-thirring1918,SSQ.1984GReGr..16..711M}
\ba
\Delta \omega_\text{Spin1} = -s\frac{12\pi \sqrt{G M} \cos I }{\left[a(1-e^2)\right]^{3/2}}
\label{SSQ.equ perihelion precession spin1}
\ea
per orbit and also there is a precession of the longitude of the ascending node
\ba
\Delta \Omega_\text{Spin1} = s\frac{4\pi \sqrt{G M}}{\left[a(1-e^2)\right]^{3/2}}
\label{SSQ.equ Delta Omega spin1}
\ea
per orbit.
Fig. \ref{SSQ.fig: jupiter Delta Omega spin1} shows the precession of the longitude of the ascending node together with the actual shift for a typical {\it Juno} orbit.

Measuring time-averaged precessions is not actually a useful strategy,
because the slightest use of spacecraft engines changes all the
Keplerian elements. But similarly to the Keplerian elements,
relativistic effects affect the observed redshift in a kick-like
manner at pericenter. Therefore, relativistic effects influence a
single pericenter passage and when the instrument is accurate enough,
they can be probed as a function of time vs. waiting for their build up
over many orbits.

\subsection{Scaling of relativistic effects}
\label{subsec scaling}


The size of the effects scale with the size of the orbit \cite{SSQ.angelilsaha2010}.
For Schwarzschild space curvature and first order spin, the respective scaling laws for the residual redshifts are
$\Delta z^\text{SS} \sim (r_G/r)^{3/2}$ and $\Delta z^\text{Spin1} \sim s (r_G/r)^{2}$ where $r_G = G M /c^2$ is the gravitational radius.
Writing distances in terms of planetary radii $r=\alpha R$, we obtain
\ba
\frac{\Delta z_1}{\Delta z_2}
= \left(\frac{s_1}{s_2}\right)^m \left(\frac{r_G^1}{r_G^2} \frac{r_2}{r_1}\right)^n
= \left(\frac{s_1}{s_2}\right)^m \left(\frac{U_1}{U_2} \frac{\alpha_2}{\alpha_1} \right)^n,
\ea
where $U_i = G M_i/(R_i c^2)$ is the gravitational potential at the surface of planet $i$ and $m=0,1$ and $n=3/2,2$ for Schwarzschild curvature and first order spin effect, respectively.
For similar orbits around different planets, i.e., $\alpha_1 = \alpha_2$ with the same eccentricity and identical Keplerian angles, this reduces to
$\Delta z_1/\Delta z_2 = (s_1/s_2)^m (U_1/U_2)^n$.
Thus, the higher the compactness $M/R$ of a planet, the higher the relativistic effect.
For frame-dragging effects, the spin parameter has also to be taken into account.


Using the expression above, we can compare the sizes of relativistic effects of orbits around the planets, the Moon and the Sun to terrestrial orbits. The ratio between the signals for similar orbits is given in Table \ref{SSQ.tab: Spin}.


\section{Planetary parameters}

The planetary parameters relevant for calculating relativistic effects
are summarised in Table~\ref{SSQ.tab: Spin}. The Moon and the Sun are
also included for comparison.

The values of the gravitational potential $U$ at the surface are
ordered as one might expect.  Jupiter with $2\times10^{-8}$ has the
highest, while for the Earth the value is 30 times smaller.

The values of the spin parameter may be surprising.  Black holes must
have $s < 1$ as is well known, but planets can have $s\gg1$. Mars
has the highest $s \sim 2090$, while Venus has the lowest $s \sim 3$,
but most planets have an $s$ with a value that is typically in the hundreds.
Incidentally, the Sun's spin parameter will be small: The Sun has a
much larger $g$ than any planet, and it spins differentially, roughly
once a month; as a result, the Sun has a much smaller $s$ than the
Earth.  The uncertainty in $s$ depends on the uncertainties in the MoI
and in the spin period.

Although neither the density profile nor internal differential
rotation can be measured directly, internal structure models provide
MoI values for the gas giants, and these are thought to be accurate to
a few percent
\cite{SSQ.helled2011constraining,SSQ.helled2011jupiter,SSQ.nettelmann2015exploration}.
The Radau-Darwin approximation \cite{SSQ.zharkov1980physics} relates
the MoI to the gravitational quadrupole $J_2$ and the ratio of
centrifugal to gravitational acceleration at the equator.  In future
it may become possible to measure planetary MOI from precession
\cite{LEMAISTRE201678}.  At present, the estimated MoI is $\sim 0.265$
for Jupiter \cite{SSQ.helled2011jupiter} and $\sim 0.220$ for Saturn
\cite{SSQ.guillot2007treatise,SSQ.helled2011constraining}.  Evidently,
Saturn is more centrally condensed than Jupiter.

The rotation period remains somewhat uncertain for all the giant planets other
than
Jupiter \cite{SSQ.helled2010uranus,SSQ.helled2009jupiter,SSQ.helled2015saturn}.
Saturn's internal rotation period is unknown to within $\sim 10$
minutes. It has been acknowledged that the rotation period is unknown
since {\it Cassini}\,'s Saturn kilometric radiation (SKR) measured a
rotation period of 10h 47m 6s \cite{SSQ.gurnett2007variable}, longer by
about eight minutes than the radio period of 10h 39m 22.4s measured by
Voyager \cite{SSQ.ingersoll1982motion}. In addition, during {\it
  Cassini}\,'s orbit around Saturn the radio period was found to be
changing with time. It then became clear that SKR measurements do not
represent the rotation period of Saturn's deep interior.  Due to the
alignment of the magnetic pole with the rotation axis, Saturn's
rotation period cannot be obtained from magnetic field measurements
\cite{SSQ.sterenborg2010can}. Theoretical efforts to infer the rotation
period \cite{SSQ.anderson2007saturn,SSQ.read2009saturn,SSQ.helled2015saturn}
indicate further sources of uncertainty. Saturn's rotation period is
thought to be between $\sim$ 10h 32m and $\sim$ 10h 47m.  For Uranus
and Neptune, the uncertainty could be as large as 4\% and 8\%,
respectively \cite{SSQ.helled2010uranus}.

A further complexity arises from the fact that the giant planets could
have non-body rotations (e.g., differential rotation on
cylinders/spheres) and/or deep winds.  However, in that case, the
deviation from a mean solid-body rotation period is expected to be
small. Future space missions to Uranus and/or Neptune, performing
accurate measurements of their gravitational fields, could be used to
determine the spin parameter of these planets.

\begin{table*}[ht]
\centering
\begin{tabular}{|l|l|l|l|l|l|l|l|}
\hline
\rowcolor[HTML]{FFFFFF} 
{\bf Object} & {$U \equiv G M/(c^2 R)$} & {$g \equiv GM/R^2$ $[m/s^2]$} & {MoI} & $s$ & {spin period [days]} & $\frac{\Delta z_\text{SS,Obj}}{\Delta z_\text{SS,Earth}}$  & $\frac{\Delta z_\text{Spin,Obj}}{\Delta z_\text{Spin,Earth}}$ \\ \hline 
Mercury & $1.00\times 10^{-10}$ & 3.7 & $0.35$ & $35.2$ & 58.65 & $5.5\times 10^{-2}$ & $9.9\times 10^{-4}$ \\ \hline
Venus & $5.98\times 10^{-10}$ & 8.9 & $0.33$ & $3.3$ & 243.02 & $8.0\times 10^{-1}$ & $3.3\times 10^{-3}$ \\ \hline
Earth & $6.95\times 10^{-10}$ & 9.8 & $0.3308$ & $738.3$ & 1.00 & $1.0$ & $1.0$ \\ \hline
Moon & $3.12\times 10^{-11}$ & 1.6 & $0.394$ & $194.8$ & 27.32 & $9.5\times 10^{-3}$ & $5.3\times 10^{-4}$ \\ \hline
Mars & $1.40\times 10^{-10}$ & 3.7 & $0.366$ & $2093.5$ & 1.02 & $9.1\times 10^{-2}$ & $1.2\times 10^{-1}$ \\ \hline
Jupiter & $2.02\times 10^{-8}$ & 25.9 & $0.265$ & $564.0$ & 0.41 & $1.6\times 10^{2}$ & $6.4\times 10^{2}$ \\ \hline
Saturn & $7.00\times 10^{-9}$ & 10.4 & $0.220$ & $988.0$ & 0.44--0.45 & $3.2\times 10^{1}$ & $1.4\times 10^{2}$ \\ \hline
Uranus & $2.52\times 10^{-9}$ & 8.9 & $0.225$ & $770.1$ & 0.67--0.76 & $6.9$ & $1.4\times 10^{1}$ \\ \hline
Neptune & $3.06\times 10^{-9}$ & 11.1 & 0.236 & 691 & 0.63--0.71 & $9.2$ & $1.8\times 10^{1}$ \\ \hline
Sun & $2.12\times 10^{-6}$ & 273.7 & $0.07$ & $0.2$ & 25.05 & $1.7\times 10^{5}$ & $2.8\times 10^{3}$ \\ \hline
\end{tabular}
\caption{Gravitational and spin parameters for the planets and the
  Moon. For the gravitational potential $U$ and acceleration $g$, values
  at the surface are given; values from orbit will be somewhat
  smaller. MoI values for the giant planets are derived using
  interior models that reproduce the gravitational fields of the
  planets \cite{SSQ.helled2015saturn}. All other quantities are derived
  using parameters provided by NASA
  [http:\//\//nssdc.gsfc.nasa.gov\//planetary\//factsheet]. 
  The two columns on the right give the ratio between the redshift signals
  of orbits around the respective object and the signals for a similar
  orbit around Earth.
  }
\label{SSQ.tab: Spin}
\end{table*}

\section{Relativistic effects for Current and Planned Missions}
\label{SSQ.sec:missions}

We now determine the effects of relativity on the redshift signal for different orbits around different planets.
In Sec. \ref{subsec Jupiter orbit} we consider a typical orbit of the {\it Juno} spacecraft around Jupiter, followed by a typical {\it Cassini} orbit around Saturn in Sec. \ref{subsec Saturn orbit}.
Finally, in Sec. \ref{subsec Earth orbit}, we discuss terrestrial orbits.

\subsection{Jupiter orbit}
\label{subsec Jupiter orbit}

On July 4, 2016, the {\it Juno} mission arrived at Jupiter and started orbiting the planet. It is
equipped to perform high precision measurements (operating at X-band and Ka-band) of its gravitational field. 
The $53$-days orbits are polar with perijove being at $\sim 1.09$ Jupiter radii and apojove at $\sim 120$ Jupiter 
radii. Such orbits provide ideal conditions for gravitational field measurements, and allow the 
spacecraft to avoid most of the Jovian radiation field. After more than four years of measurementes and $\sim 32$ orbits 
around Jupiter, {\it Juno} is planned to make one last orbit and then perform the deorbiting maneuver (see e.g., 
\cite{SSQ.matousek2007juno}).

\begin{figure}
  \centering
  \includegraphics[width=8cm]{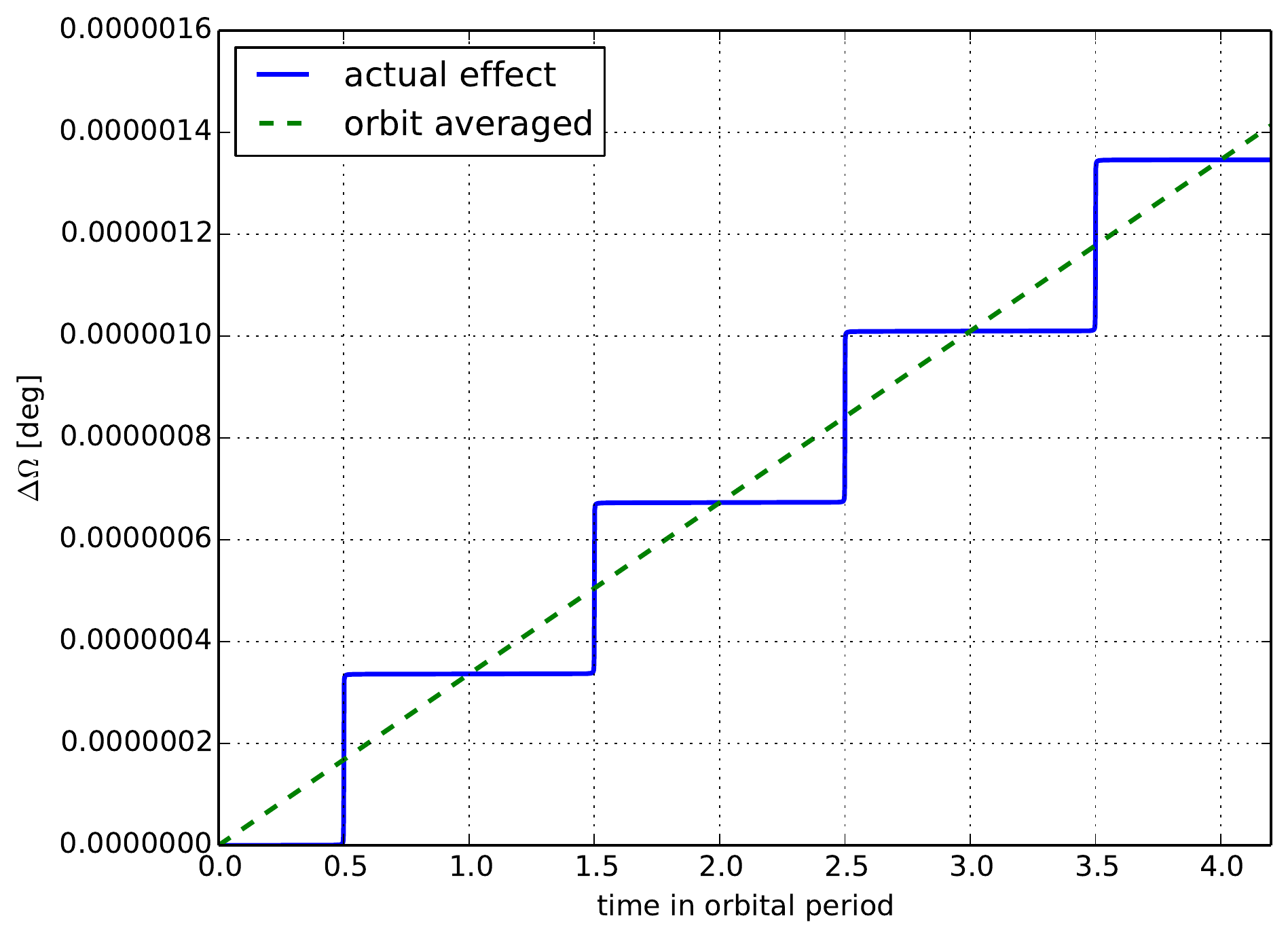}
  \caption{Change of the longitude of the ascending node
    $\Omega$ for a typical {\it Juno} orbit due to spin. The solid
    line shows the actual change of $\Omega$, while the dashed line
    represents the averaged change given by Eq. \eqref{SSQ.equ Delta Omega
      spin1}.}
\label{SSQ.fig: jupiter Delta Omega spin1}
\end{figure}

\begin{figure}
  \centering
  \includegraphics[width=8cm]{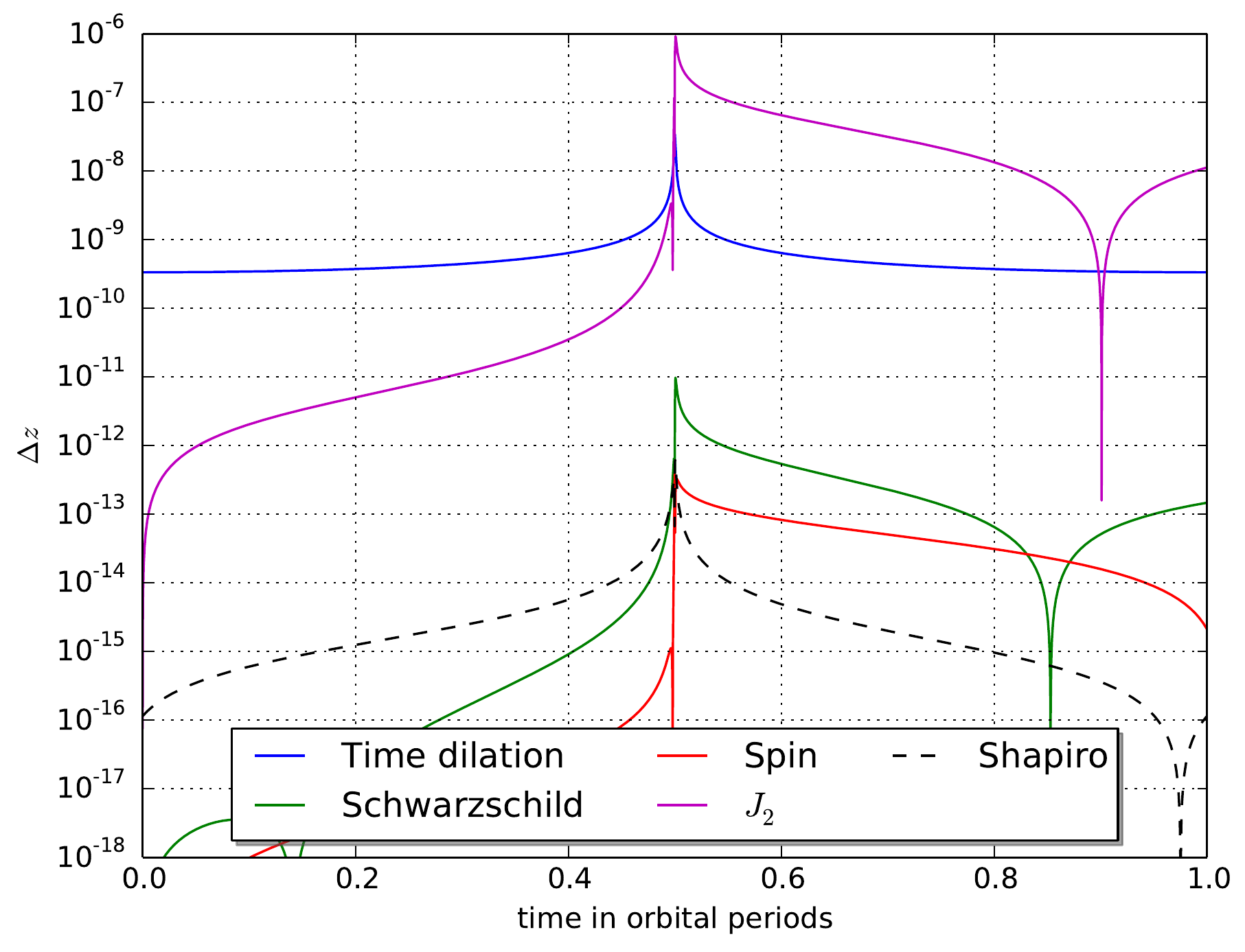}
  \caption{Higher order relativistic effects for the {\it Juno}
    orbiter. The plot shows the magnitude of the redshift signal due
    to the different relativistic effects. The parameters chosen
    correspond to a typical science orbit.  The curves change slightly
    for other orbits, however, the order of magnitude of the effects
    is the same. Also the Newtonian effect due to $J_2$ is shown.}
\label{SSQ.fig:JupiterDeltaz}
\end{figure}

\begin{figure}
  \centering
  \includegraphics[width=8cm]{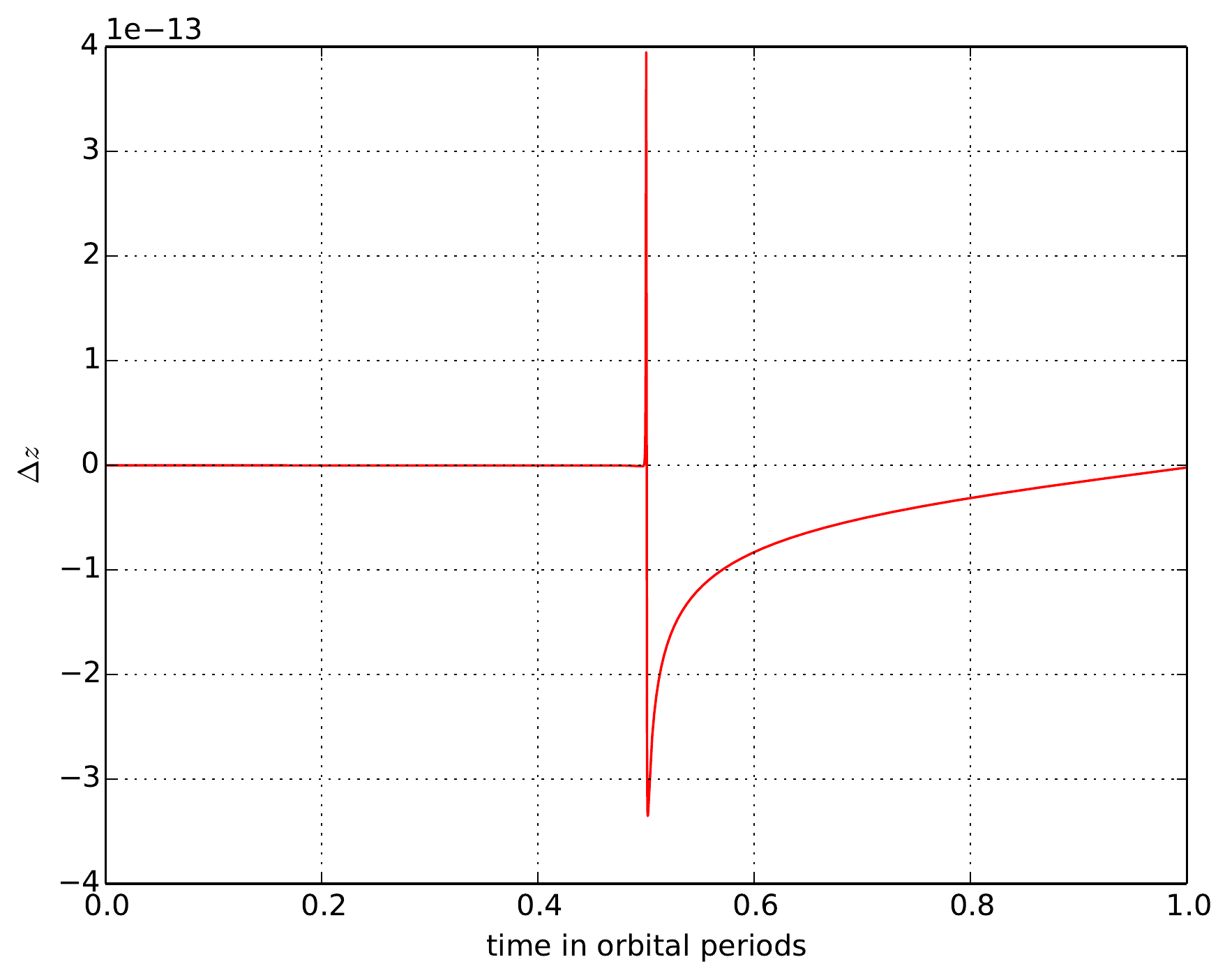}
  \includegraphics[width=8cm]{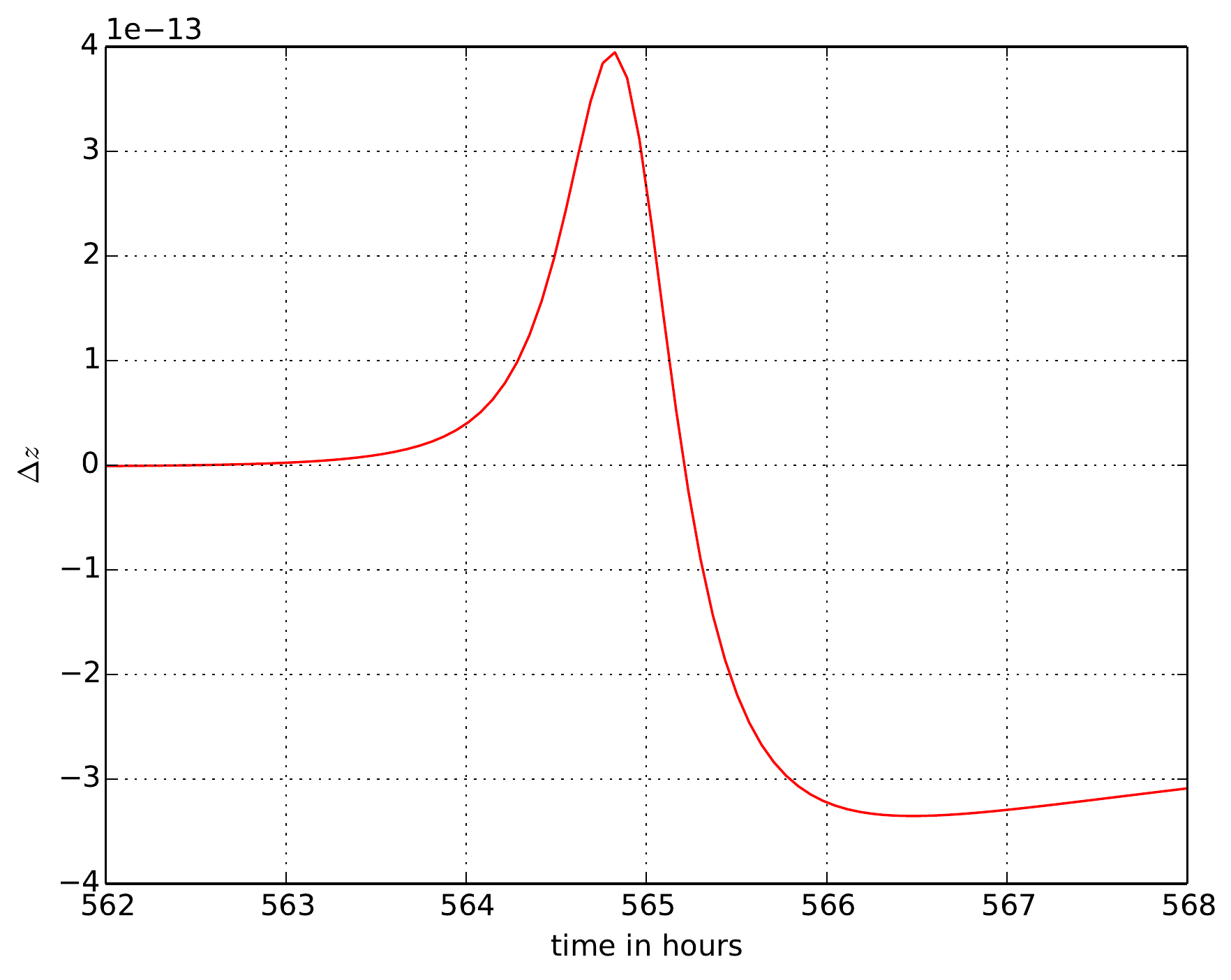}
  \caption{{\em Upper:} contribution to the redshift from
    frame-dragging by Jupiter's spin, for the same orbit as in
    Fig.~\ref{SSQ.fig:JupiterDeltaz}.  The signal peaks at the orbit
    pericenter passage. {\em Lower:}~zoom into pericenter passage.}
\label{SSQ.fig:JupiterFrameDragging}
\end{figure}

We compute the leading-order relativistic effects on the orbit of the {\it Juno} mission. They measure the precession of the orbit due to the curvature of the spacetime and contain a part that accumulates as well as a transient part, which has never been measured. The effect that occurs due to the Schwarzschild term in the Hamiltonian
produces a Mercury-like precession (solid red curve), while the other is referred to as frame-dragging due to the spin of Jupiter. Measuring the latter directly constrains the spin 
parameter of the planet, which is proportional to its moment of inertial and angular momentum. It thus reveals important information about the planet's internal 
density structure that is not necessarily identical to that contained in the gravitational moments.

The {\it Juno} orbiter has already entered a highly elliptical polar orbit around Jupiter. It is measuring deviations in the velocity of the spacecraft 
$\sim 10 \mu m$/sec $(\tau/60$ sec $)^{-1/2}$.   This corresponds to a sensitivity to redshift change of $\Delta z \sim 3 \times 10^{-14}$. 

At each pericenter passage of {\it Juno}, both the instantaneous Keplerian elements and the orientation to the observer change.
Therefore, in order to discuss relativistic effects on the basis of the {\it Juno} mission, we consider a typical orbit with average values $a = 60 \times R_\text{Jupiter}$, $e=0.981$, $\Omega=253^\circ$, $I=93.3^\circ$, $\omega=170^\circ$ and observer position $\theta_\text{obs}=92.9^\circ$ (polar angle), $\phi_\text{obs}=15.0^\circ$ (azimuthal angle). 
Fig. \ref{SSQ.fig:JupiterDeltaz} shows the characteristic redshift curves for the different effects for such a {\it Juno} orbit.
For all science orbits, the sizes of the effects, in particular of the spin effect, are similar.

Fig. \ref{SSQ.fig:JupiterFrameDragging} shows the part in the redshift due
to the presence of Jupiter's spin over one orbit.  After pericenter
passage, the relativistic and the non-relativistic orbit are out of
sync and a comparison does not make sense anymore.  The lower panel of
the figure zooms into the peak around pericenter, revealing that the
interesting time span is of order $\sim 1$ hour. This is the phase
that needs to be observed in order to seek the characteristic
imprint of frame-dragging in the redshift data.

\begin{figure}
  \centering
  \includegraphics[width=8cm]{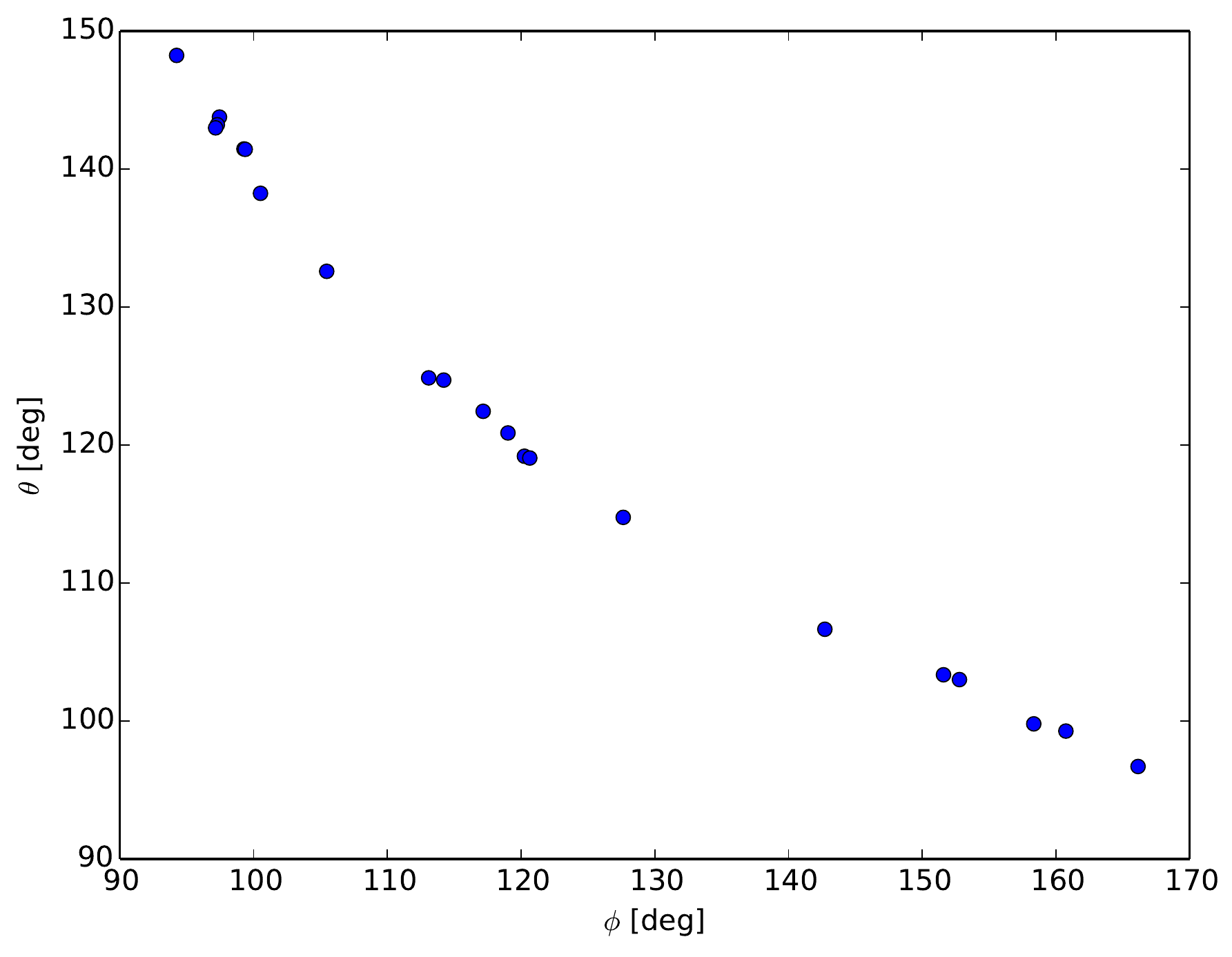}
  \caption{Orientation of the vector $\hat{r}_{\rm peri}\times\hat{b}$
    for {\it Juno} science orbits.  Here $\hat{b}$ is the line of
    sight to {\it Juno}, and $\theta,\phi$ in the Figure are with respect to
    to Jupiter's axis.  The timing signal is sensitive to the
    planetary spin projected along these various directions.}
\label{SSQ.fig:Angles}
\end{figure}

Over any one orbit, only one component of the spin vector contributes
at leading order, namely the spin component along $\hat{r}_{\rm
  peri}\times\hat{b}$ (see Eq.~\ref{SSQ.equ: Delta z Spin}).  To be
sensitive to all components of the spin, orbits with different
orientations of $\hat{r}_{\rm peri}\times\hat{b}$ are needed.
Fig.~\ref{SSQ.fig:Angles} shows the polar and azimuthal angles of this
vector for all the {\it Juno} science orbits.  The orientations are
varied, and hence {\it Juno} is sensitive to all three components of
the spin vector.



The frame-dragging effect will, moreover, be a pathfinder to measuring yet weaker effects.
The spin terms depend on the spin profile inside the planet. Measuring the spin profile would therefore play a role in 
constraining planet properties and formation models.  Future deep-space missions could enable tests of general relativity around other planets in the 
Solar System whose composition and internal structure are unknown. 

\subsection{Saturn orbit}
\label{subsec Saturn orbit}

The {\it Cassini} mission is planned to finish its exploration of the 
Saturnian system with proximal orbits around Saturn that will provide accurate measurements of the gravitational 
field of the planet. The {\it Cassini} spacecraft is planned to execute $22$ highly inclined ($63.4$ degree) 
orbits with a periapsis of $\sim 1.02$ Saturn radii \cite{SSQ.edgington2016cassini}. These proximal orbits, 
known as {\it Cassini Grand Finale}, operating at X-band, are also ideal for gravity measurements. They are 
expected to  provide range rate accuracies of $\sim 12 \mu {\rm m/sec}$ at 1000 second integration times, 
being about four times noisier than {\it Juno}. 

Both the {\it Juno} and the {\it Cassini} spacecrafts will terminate their operations by descending into the atmospheres 
of Jupiter and Saturn, respectively, and will disintegrate and burn up in order to fulfill the requirements of 
NASA's Planetary Protection Guidelines. 

{\it Cassini} has a sensitivity that is about $\Delta z \sim 10^{-13}$.
Relativistic effects peak around the pericenter with the frame-dragging effect of maximum amplitude $\sim 10^{-13}$ and the Schwarzschild curvature term
of $\sim 10^{-11}$. 
Ideally, the goal would be to resolve both the Schwarzschild and frame-dragging parts of the precession as a function of time. If they could be modeled effectively, they would less likely be drowned by Newtonian noise than a cumulative effect.

\begin{figure}
  \centering
  \includegraphics[width=8cm]{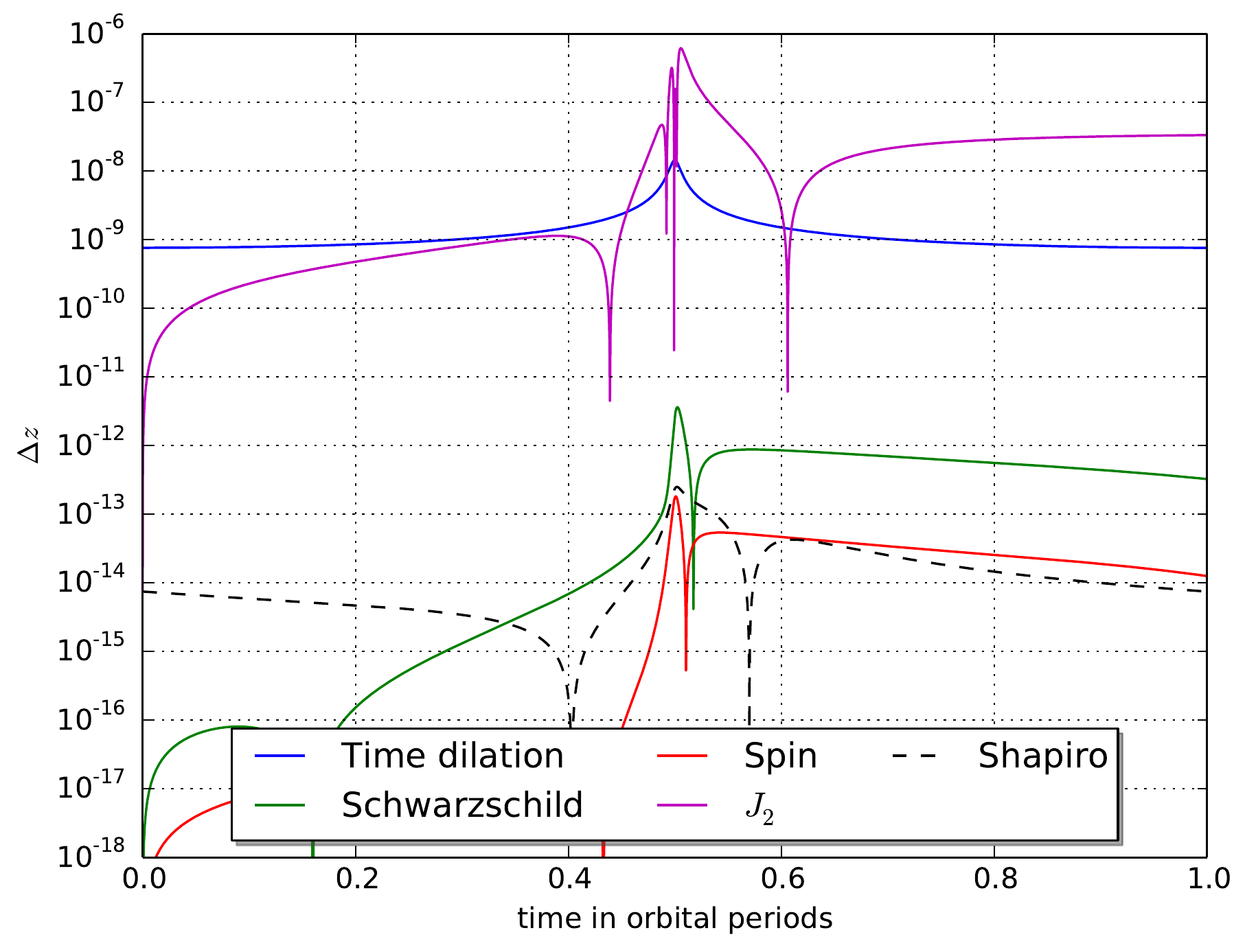}
  \caption{Higher order relativistic effects for {\it Cassini}.
  }
\label{SSQ.fig:CassiniDeltaz}
\end{figure}

Fig. \ref{SSQ.fig:CassiniDeltaz} shows the corresponding curves for a typical
{\it Cassini} orbit.  For {\it Cassini}, we chose the values $a = 10
\times R_\text{Saturn}$, $e=0.9$, $\Omega=175^\circ$, $I=62^\circ$,
$\omega=187^\circ$, $\theta_\text{obs}=63.3^\circ$ and
$\phi_\text{obs}=-5^\circ$.

\subsection{Earth orbit}
\label{subsec Earth orbit}

\begin{figure}
  \centering
  \includegraphics[width=8cm]{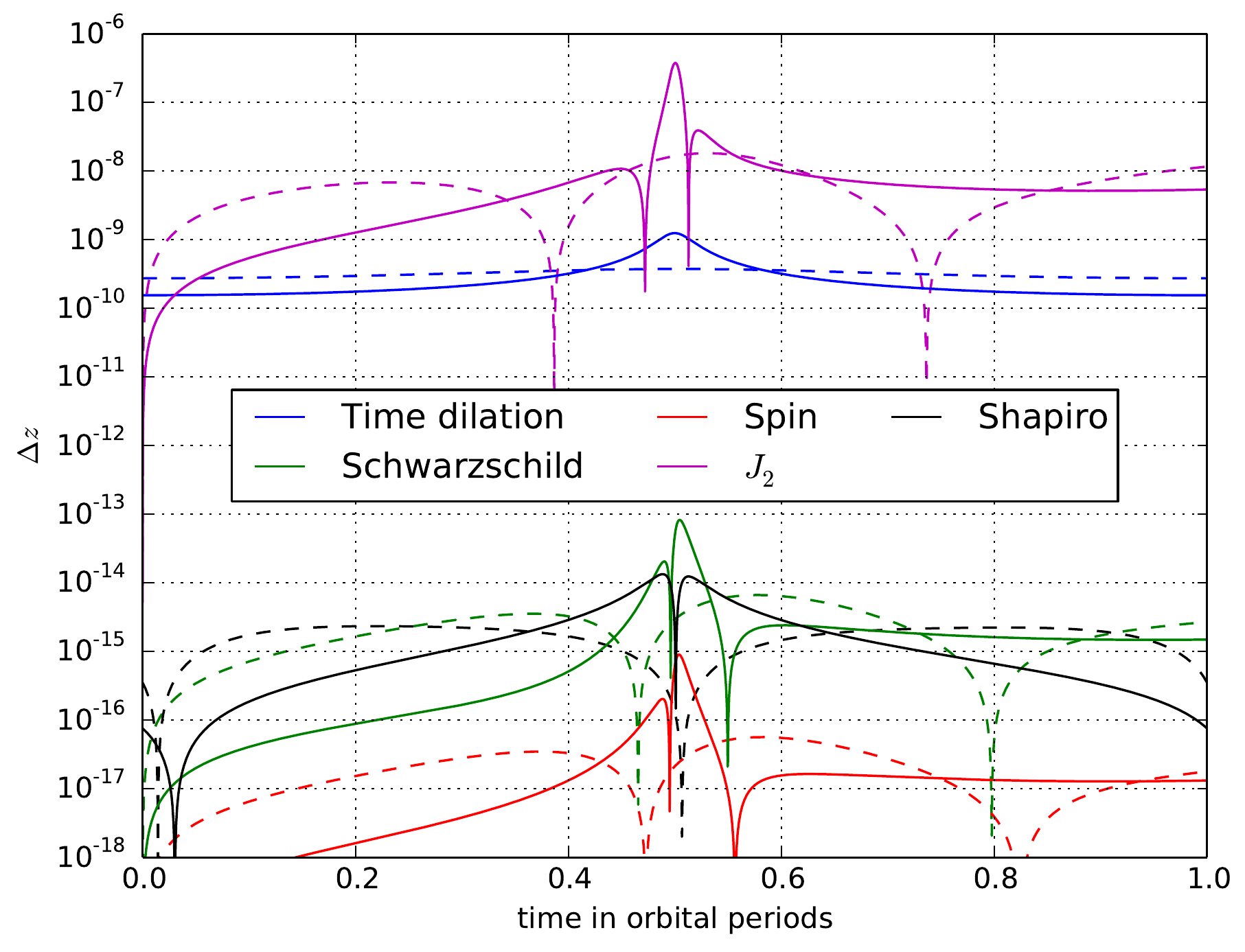}
  \caption{Redshift curves of terrestrial satellites. The dashed curves give the results for an orbit with the semi-major axis and eccentricity corresponding to the ones of the {\em Galileo} $5$ \& $6$ satellites. The solid lines give the results for a typical satellite with high eccentricity while all the other Keplerian elements and the observer's position were left the same.}
\label{SSQ.fig:EarthDeltaz}
\end{figure}

Next we discuss satellites in Earth orbit.
To illustrate the importance of eccentricity, Fig. \ref{SSQ.fig:EarthDeltaz} shows the redshift curve for a typical terrestrial satellite with a low eccentricity ($e=0.1561, a=27'977\text{km}$) as for the {\em Galileo} $5$ \& $6$ satellites and a high eccentricity ($e=0.779, a=32'090\text{km}$) orbit, while leaving all other Keplerian elements as well as the observer's position constant.
However, the actual curve depends highly on the orientation of the orbit and the position of the observer and must be computed individually for each orbit-observer-configuration.
Also, that the visibility of the satellite around pericenter might not be provided needs to be taken into account.
For the {\em Galileo} satellites, the curve would be significantly flatter - without a clear peak around pericenter due to the low eccentricity.
The only relativistic effect besides time dilation that is within the measurability range is the Schwarzschild space curvature effect.
It is expected that it will improve the currently best measurement given by Gravity Probe A \cite{Delva}.

\section{Conclusions}
\label{SSQ.conclusions}

A spinning body causes spacetime to rotate around it, thus making
nearby angular momentum vectors precess.
This had already been considered theoretically in the early days of general relativity
\cite{SSQ.lense-thirring1918}. Only in recent years, however, has the
effect entered the experimental realm
\cite{SSQ.ciufolini2004confirmation,SSQ.ciufolini2016,SSQ.everitt2011gravity}.

Frame-dragging is usually thought of as a steady precession.
For highly eccentric orbits, however, this is far from the case.
While having a minor impact along most of the orbit, frame-dragging
kicks in around pericenter.
This can be seen in Fig.~\ref{SSQ.fig: jupiter Delta Omega spin1}
which shows the change of the longitude of ascending node due to spin
for some example orbits of the {\em Juno} spacecraft.
An analogous situation applies to the S~stars in orbit
around the Galactic-center black hole \cite{SSQ.2014MNRAS.444.3780A}.
We suggest that these pericenter-kicks could provide a distinctive 
signature in timing signals obtained from spacecraft tracking.

The frame-dragging contribution to the redshift of spacecraft signals
is
\ba
\Delta z_\text{spin}
= - 2\left(\f{GM}{c^2r}\right)^2 \vec{s} \cdot (\hat{r}\times\hat{b})
\ea
(given in geometrized units as in Eq.~\ref{SSQ.equ: Delta z Spin}) where
$\hat b$ is the line of sight to the spacecraft, and $\vec s$ is the
dimensionless spin vector.  Substituting the approximation expression
\eqref{simple-spin} for the spin parameter, and assuming that the
spacecraft has a low pericenter, so that $r_\text{peri}$ is of the
same order as the planetary radius, gives
\ba
\Delta z_\text{spin} \sim \frac{GM}{c^3P}
\ea
where $P$ is the spin period.  Jupiter has $GM/c^3\sim5\,\rm nanosec$
and $P\sim10\rm\,hr$, indicating $\Delta z_\text{spin}\sim10^{-13}$.
Furthermore, as Fig.~\ref{SSQ.fig:JupiterFrameDragging} shows, the frame-dragging
signal is concentrated over a duration of two hours around the
pericenter.

In this paper we have modeled the effects of the curvature of the
spacetime on both the orbit of a spacecraft and on the electromagnetic
signals it sends to Earth. The aim is to quantify how the different
relativistic effects influence the observable redshift
signal. Geodesic equations are written in four dimensions in
Hamiltonian form.  Orbit equations for a spacecraft are a
straightforward initial-value problem, while the equations for light
signals traveling between the spacecraft and the observer form a
boundary-value problem.  Both sets of equations are solved
numerically, using extended-precision floating point arithmetic, to
compute redshift signals.  Different metric terms are turned on and
off to compare the signatures of each effect on the signal. We
particularly focus on the spin terms, for which there are good
predictions for the planets in our solar system.  The eccentricity of
the orbit can also increase the size of the terms by at least an order
of magnitude.

Figures~\ref{SSQ.fig:JupiterDeltaz}, \ref{SSQ.fig:CassiniDeltaz} and
\ref{SSQ.fig:EarthDeltaz} show example orbits of {\em Juno,} {\em
  Cassini,} and the eccentric {\em Galileo} spacecraft respectively.
They also show the effect of the quadrupole $J_2$, which is orders of
magnitude larger than the spin effect, but has a different time
dependence.  For the eccentric {\em Galileo} satellites, relativistic time
dilation reaches $\sim10^{-9}$ and is expected to be accurately
measured; the leading order effects of a Schwarzschild spacetime are
$\sim10^{-13}$ and will be challenging; spin effects are two orders of
magnitude smaller and hence unlikely to be measured.  For both {\em
  Juno\/} and {\em Cassini,} spin effects reach $\sim10^{-13}$ which is
well above timing uncertainties.

Measurability centers on whether the frame-dragging signal can be
disentangled from the much larger quadrupole and other ``foreground''
effects \cite{2011AGUFM.P41B1620F,mnras_stu2328,SERRA2016100}.
The specific and known time-dependence of the frame-dragging signal
offers some hope of doing so, but the question remains open.





\section*{Acknowledgements}

We acknowledge support from the Swiss National Science Foundation, and
thank Marzia Parisi for help with the orbits of {\em Juno} and {\em
  Cassini}. 

We also thank Luciano Iess for sharing the results of an earlier
unpublished study within the {\em Juno} mission. Their work used a
different formulation from the present one, but also concluded that
spin has an in-principle measurable effect near pericenter passages.
Furthermore, that work identified a near-degeneracy between the spin
vector and the gravitational quadrupole, leaving frame-dragging
measurable by {\em Juno} only if the spin axis is independently
precisely constrained.

\bibliography{framedragging.bib}

\end{document}